\documentclass[9pt,twocolumn,twoside]{osajnl}
\journal{ol}
\setboolean{shortarticle}{true}
\usepackage{graphicx}
\usepackage{txfonts}
\usepackage[english]{babel}
\newlength{\figsize}
\newlength{\subfigsize}
\begin{document}

\title{Optical Kerr nonlinearity of dielectric nanohole array metasurface in proximity to anapole state}

%\maketitle

\author{Andrey V. Panov}

%\author{Andrey V. Panov\authormark{1}}

% \ead{andrej.panov@gmail.com}
% \authorrunning{A. Panov}
% \email{Electronic mail: andrej.panov@gmail.com}
% \mail{\email{andrej.panov@gmail.com}}
% \email{andrej.panov@gmail.com}
% \email{Electronic mail: panov@iacp.dvo.ru}
% \address[1]{
% \affiliation{
%\institute{
 \affil[1]{
% \address{
% \begin{affiliations}
 
Institute of Automation and Control Processes,
Far Eastern Branch of Russian Academy of Sciences,
5, Radio st., Vladivostok, 690041, Russia}
\affil[*]{Corresponding author: panov@iacp.dvo.ru}
%\email{\authormark{*}andrej.panov@gmail.com}
\dates{Received 7 April 2022; revised 6 May 2022; accepted 10 May 2022; published 26 May 2022}
%\ociscodes{(190.4400)   Nonlinear optics, materials; (260.2065)   Effective medium theory; (160.4330)   Nonlinear optical materials}
\setboolean{displaycopyright}{true}
\setcounter{page}{2866}
\renewcommand*{\journalname}{Vol. 47, No. 11 / 1 June 2022 / Optics Letters}
\doi{\url{https://doi.org/10.1364/OL.459989}}
%\keywords{Anapole state, Optical Kerr nonlinearity, Dielectric metasurface}

\begin{abstract}
% \abstract{
Nowadays, metasurfaces have attracted a great deal of attention from researchers due to prominent optical properties.
In particular, the metasurfaces may consist of structures possessing optical anapole resonances with strong field confinement and substantially suppressed scattering. 
As a result, such nanostructures display enhanced nonlinear optical properties.
In this paper by means of three-dimensional finite-difference time-domain simulations, the ability of anapole modes in high-index dielectric metasurfaces with circular nanopores is shown. 
In the vicinity of the anapole state, the effective optical Kerr nonlinearity increases by orders of magnitude. 
Simultaneously, the optical transmission of the metasurface can reach high values up to unity.
% }
\end{abstract}

%\begin{document}

\maketitle
% \section{Introduction}

In recent years, there has been considerable interest in the study of optical anapoles. 
In nanostructures, anapole modes arising from destructive interference of the electric and toroidal dipole resonant modes possess strong field enhancement and reduced scattering properties. 
The suppressed scattering prefers these modes to Mie-type resonances which are also observed in nanostructures but have larger radiation losses. 
As a consequence of the electric field confinement, the extraordinary enlargement of the optical nonlinearity is revealed. 
Initially it was demonstrated that the second- \cite{Rocco18,Timofeeva18} or third-harmonic \cite{Grinblat16} conversion efficiencies of a dielectric nanodisks at the anapole state are enhanced by the orders of magnitude of those of unstructured materials. 
As shown in Ref.~\cite{Grinblat20}, the optical Kerr effect (OKE) and two-photon absorption in a crystalline gallium phosphide (GaP) nanodisks at the anapole excitation allow to achieve efficient ultrafast all-optical modulation. 
Zhang et al. \cite{Zhang20} observed anapole-mediated photothermal nonlinearity in silicon with three orders of magnitude enhancement as compared with that of bulk Si.

For dielectric materials, optical anapoles were seen in disks \cite{Miroshnichenko15}, core-shell cylinders \cite{Song20}, ellipsoids \cite{Ospanova21}, spheres \cite{Parker20}, nanocuboids 
\cite{%Sun18a,
Algorri19}, $\Phi$-shaped dielectric nanostructures \cite{Wu20}, grouped nanohole arrays \cite{Ospanova18a}. 
The fabrication of holey metasurfaces seems to have some benefits over other shapes. %\cite{Tasolamprou17}
Moreover, nanohole arrays can be utilized for freestanding dielectric metasurfaces (membranes) which are elaborated now for visible and near-infrared ranges \cite{Karvounis18,Lim21}. 
As such the occurrence of the toroidal field distribution viewed in grouped nanohole arrays does not guarantee a high concentration of the field and hence large optical nonlinearity.
The actual enhancement of the optical nonlinearity by the nanostructure should be checked.

In the present work, it is shown by numeric simulations that a simple square lattice array of nanoholes inside a high refractive index dielectric slab may have anapole states with the optical nonlinearity enlarged by orders of magnitude.
This is done with the help of a numerical technique for retrieving the effective Kerr nonlinearity of nanocomposites \cite{Panov18}. 
Aforementioned method is based on three-dimensional finite-difference time-domain (FDTD) simulations of light propagation through a sample with optical nonlinearity.
A change of optical phase of the Gaussian beam transmitted through the sample at different light intensities $I$ allows one to estimate the nonlinear refractive index resulting from  optical Kerr effect and defined by
\[
 n=n_0+n_2 I,
\]
where $n_0$ is the linear refractive index, $n_2$ is the second-order nonlinear refractive index.
The nonlinear index of refraction is calculated at several locations on the transmitted beam axis that permits one to estimate the mean value and the standard deviation of $n_2$.

% \section{FDTD simulation details}

% The three-dimensional simulations of the Gaussian beam propagation through the nonlinear medium are performed with the Massachusetts Institute of Technology (MIT) Electromagnetic Equation Propagation (MEEP) FDTD solver \cite{OskooiRo10}.
% The modeling of light scattering by the lattice elements is done with openEMS (Open Electromagnetic Field Solver) \cite{openEMS} and further processed with MENP (an open-source MATLAB implementation of multipole expansion for nanophotonics) \cite{Hinamoto21}.

In this study, the modeling of high index medium for the visible range (wavelength $\lambda=532$~nm unless otherwise specified in the context) is done for GaP with
the linear refractive index of the metasurface material $n_{0\,\mathrm{in}}$ being equal to 3.49 \cite{Aspnes83} with negligible  extinction coefficient.
 The value of second-order nonlinear refractive index $n_{2\,\mathrm{bulk}}$ of GaP is set to $6.5\times10^{-17}$~m$^2$/W which is based on the measurements of the third-order optical susceptibility in the visible range \cite{Kuhl85}.
% The size of the FDTD computational domain for simulations in the visible range is $2.8\times 2.8\times 15$~$\mu$m, the space resolution of the simulations is 3.3~nm. 
For the infrared range ($\lambda=1034$~nm), silicon ($n_{0\,\mathrm{in}}=3.56$) is selected as it has low absorption and substantial optical nonlinearity at this wavelength. %\cite{Schinke15}
 The value of second-order nonlinear refractive index for Si at this range $n_{2\,\mathrm{bulk}}\approx4\times10^{-17}$~m$^2$/W \cite{Bristow07}.
% The size of the computational domain for simulations at this wavelength is $4\times 4\times 30$~$\mu$m with the resolution of  4~nm.
The surrounding medium is vacuum.
Details of the FDTD simulations are provided in Supplement 1.
Figure~\ref{nldisk_pores} shows a schematic representation of the nanohole array metasurface.
Here $r$ is the radius of the circular nanohole, $a$ is the lattice constant, $h$ is the thickness of the metasurface.
The simulated linearly polarized Gaussian beam falls perpendicularly on the metasurface. 
Except where otherwise noted, the electric field of the incident beam is polarized along the $x$-axis.

\begin{figure}
{\centering\includegraphics[width=\figsize]{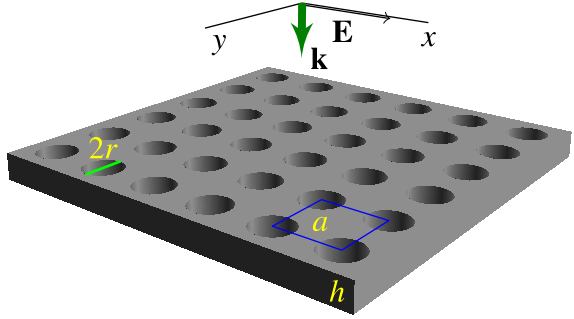}\par} 
\caption{\label{nldisk_pores}  
Schematic of the simulated metasurface comprising a square lattice array of nanoholes in a high refractive index slab. 
The Gaussian beam is incident normally on the metasurface. }
\end{figure} 

% \section{Results and discussion}

As previously shown by numerical simulations \cite{Panov20}, the anapole state appears in GaP disks at $\lambda=532$~nm with a height of 50 nm and a radius of approximately 126~nm. 
After modeling in this investigation, it is found that the GaP nanohole array with $h=100$~nm, $a=250$~nm and $r\approx80{-}90$~nm excited by a linearly polarized Gaussian beam exhibits a double toroidal distribution of electric field energy. 
The time-averaged electric $|E|^2$ and magnetic $|H|^2$ energy distributions at cross-section of the metasurface at $h/2$ are illustrated in Fig.~\ref{ener_dist_GaP_361_lattcylhol}. 

\begin{figure}[tb!]
{\centering
\includegraphics[width=\subfigsize]{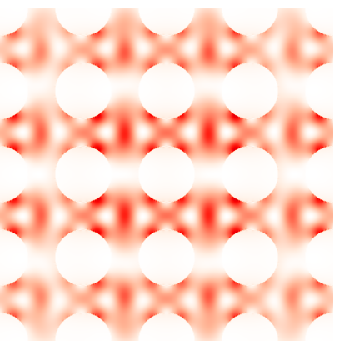}\hspace{1em}
\includegraphics[width=\subfigsize]{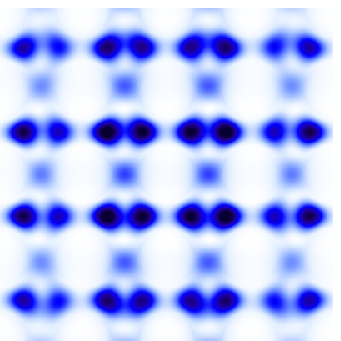}\par}
\caption{\label{ener_dist_GaP_361_lattcylhol} 
Time-averaged distributions of electric $|E|^2$ (left part, red color) and magnetic $|H|^2$ (right part, blue color) energy densities in the GaP nanohole array at the anapole mode ($a=250$~nm, $h=100$~nm, $r=84$~nm, $\lambda=532$~nm).
The distributions are calculated within the plane $h/2$.
The incident Gaussian beam is polarized along the vertical direction.}
\end{figure}

\begin{figure}[tb!]
{\centering
\includegraphics[width=\figsize]{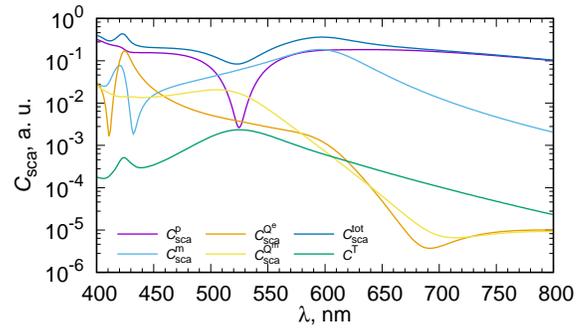}
\par} 
\caption{\label{Csca_exact_toroidalz_log_r084}
Scattering cross section spectra for the multipole contributions (electric dipole $C_\mathrm{sca}^\mathrm{p}$, magnetic dipole $C_\mathrm{sca}^\mathrm{m}$, electric quadrupole $C_\mathrm{sca}^\mathrm{Q^e}$, magnetic quadrupole $C_\mathrm{sca}^\mathrm{Q^m}$), their sum $C_\mathrm{sca}^\mathrm{tot}$ and the intensity of the electric dipole toroidal moment $C^\mathrm{T}$ for the GaP lattice element with $a=250$~nm, $h=100$~nm, $r=84$~nm. 
Refractive index $n_{0\,\mathrm{in}}$ is assumed to be constant over the whole wavelength range.
}
\end{figure}

Figure~\ref{Csca_exact_toroidalz_log_r084} represents a multipole decomposition of scattering cross sections for a lattice element delineated by the blue-lined square in Fig.~\ref{nldisk_pores}. 
The electric and magnetic multipole scattering cross sections are calculated here with the exact formulas as defined in Ref.~\cite{Alaee18}. 
The dipole electric toroidal moment $\mathbf{T}$ is described by the intensity
\[
C^\mathrm{T}=
\frac{k^4}{6\pi\varepsilon_0^2 |\mathbf{E}_{\mathrm{in}}|^2}\left|
%\sum_\alpha=1^3
\mathbf{T}
\right|^2
,
\]
\[
\mathbf{T}=%\frac{k^2}{2}
% \left|
\int d\mathbf{r}\left\lbrace 3(\mathbf{r}\cdot\mathbf{J})
\mathbf{r}
%r_\alpha
-r^2 
\mathbf{J}
% J_\alpha
\right\rbrace\frac{j_2(kr)}{2r^2}%{(kr)^2}
% \right|^2
, 
\]
where $|\mathbf{E}_{\mathrm{in}}|$ is the electric field amplitude of the incident wave, $k$ is the wavenumber in vacuum, 
$\varepsilon_0$ is the vacuum permittivity, $\mathbf{J}$ is the induced electric current density, $j_2(kr)$ is the spherical Bessel function. 
Being widely used elsewhere, the expression for $\mathbf{T}$ in the long-wave approximation is not utilized here as it does not work well for the objects in the anapole state.
It should be noted that $C^\mathrm{T}$ makes no explicit contribution to the total scattering cross section $C_\mathrm{sca}^\mathrm{tot}$.
As can be seen from Fig.~\ref{Csca_exact_toroidalz_log_r084}, the total scattering cross section $C_\mathrm{sca}^\mathrm{tot}$ and the electric dipole cross section $C_\mathrm{sca}^\mathrm{p}$ have minima near the wavelength of interest 532~nm, while the toroidal dipole intensity $C^\mathrm{T}$ reaches a maximum there.

\begin{figure}[tb!]
{\centering
\includegraphics[width=\figsize]{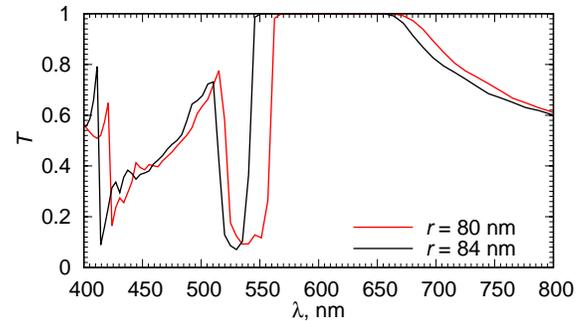}
\par} 
\caption{\label{Tspectr_GaP_r0.84}
Transmission spectra for the GaP nanohole array with $a=250$~nm, $h=100$~nm in the vicinity of the anapole state. 
Refractive index $n_{0\,\mathrm{in}}$ is assumed to be constant over the whole wavelength range.
}
\end{figure}

\begin{figure}[tb!]
{\centering
\includegraphics[width=\figsize]{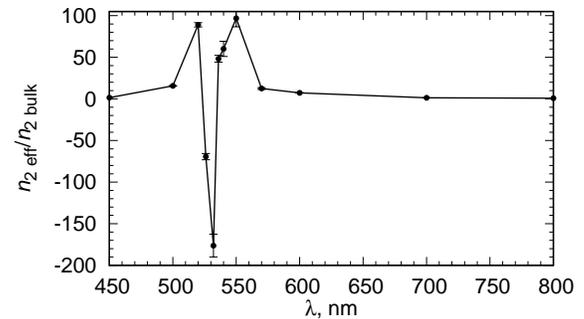}
\par} 
\caption{\label{n2_lam_GaP_361_lattcylhol}
Enhancement of the effective second order refractive index of the GaP nanohole array with $a=250$~nm, $h=100$~nm, $r=80$~nm as a function of the wavelength.
}
\end{figure}

The transmission spectra of the GaP nanohole arrays with $a=250$~nm, $h=100$~nm, $r=80$~nm and $r=84$~nm are depicted in Fig~\ref{Tspectr_GaP_r0.84}. 
Near the anapole state the transmission becomes close to unity and the metasurface exhibits anapole-induced transparency.
It should be noticed the dip in proximity to the anapole state. 
This dip corresponds to a spike in the scattering and the negative second order refractive index (Fig.~\ref{n2_lam_GaP_361_lattcylhol}). 

Further, the effective OKE of the modeled nanohole array is investigated. 
Figure~\ref{n2_lam_GaP_361_lattcylhol} demonstrates the spectral dependence of the enhancement of the effective OKE of the GaP nanohole array with $a=250$~nm, $h=100$~nm, $r=80$~nm. 
This enhancement is defined as the ratio of the effective second order refractive index of the metasurface $n_{2\mathrm{eff}}$ to the bulk $n_2$ of its material.
In the vicinity of the anapole state, the effective second order refractive index of the nanohole array is larger by two orders of magnitude than that of the unstructured material.
Utilizing the value of $n_{2\,\mathrm{bulk}}=6.5\times10^{-17}$~m$^2$/W, %of GaP from Ref.~\cite{Panov20}, 
the magnitude of $n_{2\mathrm{eff}}\approx -1.1\times10^{-14}$~m$^2$/W at $\lambda=532$~nm.
The value of $|n_{2\mathrm{eff}}|$ of the nanohole array is an order of magnitude lower than that of the ordered array of the GaP nanodisks in Ref.~\cite{Panov20}.
In fact, the existence of the toroidal mode does not assure the field confinement and the enhancement of the optical nonlinearity.
Using the technique of the present work, it is possible to show that the grouped nanohole array in the Si slab with $h=100$~nm, $r=22.5$~nm from Ref.~\cite{Ospanova18a} gives $n_{2\mathrm{eff}}\approx 2 n_{2\mathrm{bulk}}$, i.e. their structure does not display the valuable enlargement of the optical nonlinearity.

In proximity to the anapole state, the sign of  $n_{2\mathrm{eff}}$ changes to negative. 
The same effect was calculated before for the disks at the anapole state \cite{Panov20} and for the spheres at the Mie resonances \cite{Panov19}.
The similar phenomenon was observed experimentally for the effective nonlinear dielectric function of the Ag, Cu and Au nanoparticles in silica glass in the vicinity of a localized surface-plasmon resonance \cite{Takeda07,Takeda08,Sato14}.
This effect in plasmonic nanocomposites was interpreted with the help of the local field enhancement inside the particle within the framework of Maxwell-Garnett effective medium model.
Similar effective medium approximations are not appropriate for the high index nanostructures due to the large sizes of their elements.
It should be emphasized that the dip in the $n_{2\mathrm{eff}}$ corresponds to the plunge in the transmission (Fig~\ref{Tspectr_GaP_r0.84}).
This effect demonstrates that the anapole mode does not prevail in this region.

\begin{figure}[tb!]
{\centering
\includegraphics[width=\figsize]{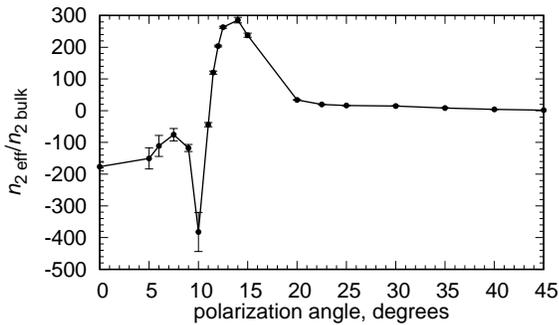}
\par} 
\caption{\label{n2_yrot_GaP_361_lattcylhol}
Enhancement of the effective second order refractive index of the GaP nanohole array with $a=250$~nm, $h=100$~nm, $r=80$~nm as a function of the angle between polarization direction and  $x$-axis.
}
\end{figure} 

The dependence of $n_{2\mathrm{eff}}/n_{2\mathrm{bulk}}$ for GaP nanohole array with  $a=250$~nm, $h=100$~nm, $r=80$~nm with both signs of $n_{2\mathrm{eff}}$ on the angle between the electric field polarization direction and $x$-axis is displayed in Fig.~\ref{n2_yrot_GaP_361_lattcylhol}. 
Because of the symmetry, the angle magnitude is limited by $45^\circ$.
This dependence is not monotonic and the sign of $n_{2\mathrm{eff}}$ changes since the intensity of the toroidal mode varies with the angle. 
%Fig. \ref{ener_dist_GaP_361_lattcylhol_rot20} illustrates 
Fig.~S3 (Supplement 1) illustrates 
the distortion of the field pattern due to the rotation of the polarization direction.

% \begin{figure}[tb!]
% {\centering
% \includegraphics[width=\subfigsize]{E_D_sz28sy14_res40_t10.64_eps1_conc0.674794_epsin12.203_amp1_r0.8.eps}\hspace{1em}
% \includegraphics[width=\subfigsize]{E_H_sz28sy14_res40_t10.64_eps1_conc0.674794_epsin12.203_amp1_r0.8.eps}\par}
% \caption{\label{ener_dist_GaP_361_lattcylhol_rot20} 
% Time-averaged distributions of electric $|E|^2$ (left part, red color) and magnetic $|H|^2$ (right part, blue color) energy densities in the GaP nanohole array at the anapole mode ($a=250$~nm, $h=100$~nm, $r=80$~nm, $\lambda=532$~nm).
% The distributions are calculated within the plane $h/2$.
% The incident Gaussian beam is polarized along the vertical direction. The nanohole array is rotated by $20^\circ$ to the electric field polarization direction.}
% \end{figure} 

\begin{figure}[tb!]
{\centering
\includegraphics[width=\figsize]{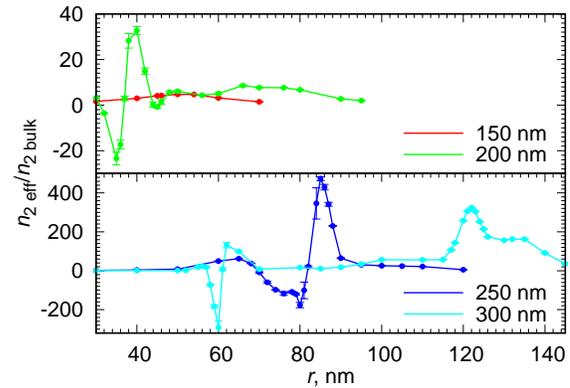}
\par} 
% \includegraphics[width=\figsize]{n2_r_a_lattcylhol_GaP_1_n2_lin.eps}
% \par} 
% {\centering
% \includegraphics[width=\figsize]{n2_r_a_lattcylhol_GaP_2_n2_lin.eps}
% \par} 
\caption{\label{n2_r_a_GaP_361_lattcylhol}
Enhancement of the effective second order refractive index of the GaP nanohole arrays with $h=100$~nm as functions of the nanohole radius $r$ and the lattice parameter $a$.
}
\end{figure} 

Next, the influence of the geometric parameters of the GaP nanohole array metasurface on the effective OKE is studied. 
The dependence of the effective second-order refractive index on the lattice parameter $a$ and the hole radius $r$ is depicted in Fig.~\ref{n2_r_a_GaP_361_lattcylhol}. 
A valuable enhancement of $n_{2\mathrm{eff}}$ appears for $a=200~$~nm, then it approaches a maximum for $a=250$~nm.
The value of $r$ for the anapole state gradually increases as $a$ rises.
The lattice with $a=300$~nm has the anapole mode near $r=120$~nm.
The intensity of the toroidal mode $C^\mathrm{T}$ is reduced as compared to that of the metasurface with $a=250$~nm (Supplement 1 Fig.~S4). 
Another peak of $n_{2\mathrm{eff}}$ for $a=250$~nm near $r=60$~nm is caused by the electric dipole resonance in the regions between the adjacent nanoholes.

\begin{figure}[tb!]
{\centering
\includegraphics[width=\figsize]{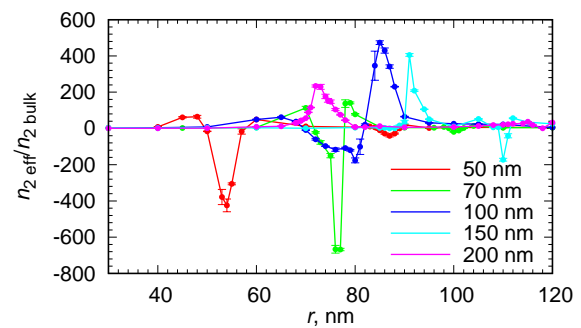}
\par} 
\caption{\label{n2_r_h_GaP_361_lattcylhol}
Enhancement of the effective second order refractive index of the GaP nanohole arrays with $a=250$~nm as functions of the nanohole radius $r$ and the metasurface thickness $h$.
}
\end{figure} 

Figure~\ref{n2_r_h_GaP_361_lattcylhol} displays the dependence of the effective nonlinear Kerr coefficient of the GaP nanohole arrays with the fixed lattice parameter $a=250$~nm on their thickness and the hole radius $r$. 
For $h<100$~nm the anapole state occurs at the lower values of $r$.
The metasurface with $h=150$~nm possesses the anapole state approximately at $r=90$~nm.
Thicker metasurface with $h=200$~nm has complicated resonance structure without the anapole modes.
Thinner nanostructures with dips of $n_{2\mathrm{eff}}$ have lower transmission (Fig. S5 in Supplement 1) which demonstrates that the anapole mode is reduced as compared to $h=100$~nm.
Thus, by adjusting geometric characteristics it is possible to achieve the anapole state with the large enhancement of the optical nonlinearity and high transmission.

\begin{figure}[tb!]
{\centering
\includegraphics[width=\figsize]{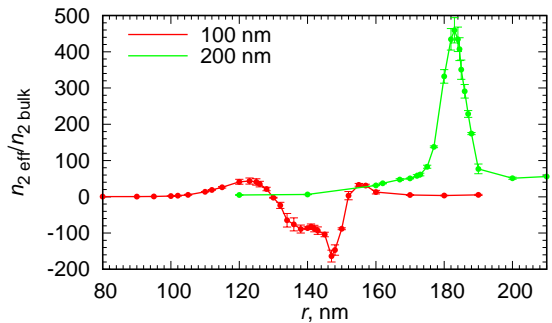}
\par} 
\caption{\label{n2_r_h_Si_169_lattcylhol}
Enhancement of the effective second order refractive index of the GaP nanohole arrays with $a=500$~nm as functions of the nanohole radius $r$ and the metasurface thickness $h$ for $\lambda=1034$~nm.
}
\end{figure} 

\begin{figure}[tb!]
{\centering
\includegraphics[width=\subfigsize]{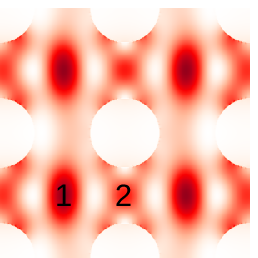}\hspace{1em}
\includegraphics[width=\subfigsize]{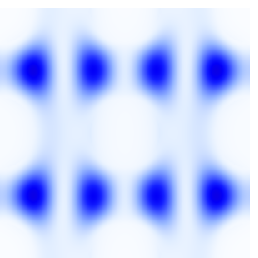}\par}
\caption{\label{ener_dist_Si_169_lattcylhol} 
Time-averaged distributions of electric $|E|^2$ (left part, red color) and magnetic $|H|^2$ (right part, blue color) energy densities in the Si nanohole array at the anapole mode ($a=500$~nm, $h=100$~nm, $r=138$~nm, $\lambda=1034$~nm).
The distributions are calculated within the plane $h/2$.
The incident Gaussian beam is polarized along the vertical direction.}
\end{figure}

Another high index material---silicon which has low absorption and substantial nonlinearity in the infrared range also exhibits spikes of the effective optical Kerr nonlinearity at resonances (Fig.~\ref{n2_r_h_Si_169_lattcylhol}). 
The geometric parameters for the resonances are approximately twice that of GaP due to the larger wavelength while the refractive index of Si is close to that of gallium phosphide.
The enhancement of the intensity of the electric dipole toroidal moment $C^\mathrm{T}$ is less pronounced for silicon in the infrared range (Supplement 1 Fig.~S6%\ref{Csca_Si_exact_toroidalz_log}
).
There are two regions of the electric field confinement for the Si nanohole arrays with  $h=100$~nm:
the first in the center of the area selected by the blue-lined square in Fig.~\ref{nldisk_pores} (denoted by 1 in Fig.~\ref{ener_dist_Si_169_lattcylhol}) and the second between two adjacent pores (indicated by 2 Fig.~\ref{ener_dist_Si_169_lattcylhol}).
The interplay between these modes defines  the behavior of  $n_{2\mathrm{eff}}$ within the range of $r=120{-}150$~nm in Fig.~\ref{n2_r_h_Si_169_lattcylhol}.
The transmission coefficient $T$ for the Si metasurface is lower for the modeled parameters at the wavelength of interest (Supplement 1 Fig.~S7%\ref{Tspectr_Si}
).
For the Si nanohole arrays with $h=100$~nm and $r=123$~nm, $T$ has a maximum near $\lambda=1034$~nm due to the first region of the electric energy confinement.
The second region causes a dip in the transmission for $h=100$~nm and $r=147$~nm.
As for Mie scattering, the scattering cross section for the anapole modes is expected to depend on the relation of geometric characteristics, $\lambda$ and $n_{0\,\mathrm{in}}$.
So the optimal parameters for the Si nanohole array are yet to be determined.

% \begin{figure}[tb!]
% {\centering
% \includegraphics[width=\figsize]{Csca_Si_exact_toroidalz_log_r180.eps}
% \par} 
% \caption{\label{Csca_Si_exact_toroidalz_log}
% Scattering cross section spectra for the multipole contributions (electric dipole $C_\mathrm{sca}^\mathrm{p}$, magnetic dipole $C_\mathrm{sca}^\mathrm{m}$, electric quadrupole $C_\mathrm{sca}^\mathrm{Q^e}$, magnetic quadrupole $C_\mathrm{sca}^\mathrm{Q^m}$), their sum $C_\mathrm{sca}^\mathrm{tot}$ and the intensity of the electric dipole toroidal moment $C^\mathrm{T}$ for the Si lattice element with $a=500$~nm, $h=200$~nm, $r=180$~nm. 
% Refractive index $n_{0\,\mathrm{in}}$ is assumed to be constant over the whole wavelength range.
% }
% \end{figure} 

% \begin{figure}[tb!]
% {\centering
% \includegraphics[width=\figsize]{T_lattcylhol_Si.eps}
% \par} 
% \caption{\label{Tspectr_Si}
% Transmission spectrum for the Si nanohole array with $a=500$~nm. Refractive index $n_{0\,\mathrm{in}}$ is assumed to be constant over the whole wavelength range.
% }
% \end{figure} 

Furthermore, it is of importance to check the influence of substrate on the field distribution inside the nanohole array.
The modeling of the GaP metasurface on the substrate with $n=1.5$ shows that the field pattern of Fig.~\ref{ener_dist_GaP_361_lattcylhol} is mainly preserved but the maximum intensity of the electric field energy $|E|^2$ at $h/2$ is reduced by 30~\%.
Thereby, the high index nanohole arrays on the substrate may be utilized for the field confinement and the enlargement of the optical nonlinearity.

% \section*{Conclusions}

In summary, it is shown that the anapole states for the high index metasurface comprising the circular nanohole array are possible.
The transmission of the nanohole array varies greatly near the anapole state: from the large scattering and the low transmission to the full transparency.
The real part of the effective Kerr nonlinear refractive index of the metasurface is estimated.
In proximity to the anapole state, there are two orders of magnitude enhancement of the second order nonlinear refractive index compared to unpatterned material.
As follows, there is a potential for achieving both the spike in the optical nonlinearity and the high transmission by the selection of the parameters of the high index nanohole array.

% \subsubsection*{Acknowledgements} 
\paragraph{Acknowledgement.} 
%\begin{acknowledgement}
 The results were obtained with the use of IACP FEB RAS Shared Resource Center ``Far Eastern Computing Resource'' equipment (https://www.cc.dvo.ru).

%\subsubsection*{Disclosures} 
\paragraph{Disclosures.}
The author declares no conflicts of interest.

\paragraph{Supplemental document.}
See Supplement 1 for supporting content.

\paragraph{Data availability.}
Data underlying the results presented in this paper are not publicly available at this time but may be obtained from the authors upon reasonable request.

\bibliography{nlphase}

% \bibliographyfullrefs{nlphase}

\clearpage
%\onecolumn
\setcounter{page}{1}
\setcounter{figure}{0}
\renewcommand*{\journalname}{Supplement 1}
\renewcommand{\thefigure}{S\arabic{figure}}
\renewcommand{\thepage}{S\arabic{page}}

%\begin{figure*}[t!]
%\section*{Supplement 1}

%\begin{abstract}
{\bfseries
\noindent
This document provides supplementary information for ``Optical Kerr nonlinearity of dielectric nanohole array metasurface at anapole state''. 
Packages used for modeling are listed.
A description of the geometric parameters for FDTD simulations are given.
Additional graphs for the nanohole nanostructures are shown.

 }
%\end{abstract}

%\end{figure*}

%\twocolumn
%\maketitle

\section*{FDTD simulation details}

The three-dimensional simulations of the Gaussian beam propagation through the nonlinear medium are performed with the Massachusetts Institute of Technology (MIT) Electromagnetic Equation Propagation (MEEP) FDTD solver \cite{OskooiRo10}.
The size of the FDTD computational domain for simulations in the visible range is $2.8\times 2.8\times 15$~$\mu$m, the space resolution of the simulations is 3.3~nm. 
The size of the computational domain for simulations at this wavelength is $4\times 4\times 30$~$\mu$m with the resolution of  4~nm.
The modeling of light scattering by the lattice elements is done with openEMS (Open Electromagnetic Field Solver) \cite{openEMS} and further processed with MENP (an open-source MATLAB implementation of multipole expansion for nanophotonics) \cite{Hinamoto21}.
The space resolution for the modeling with openEMS is 2.5 nm.
The linear refractive index $n_{0\,\mathrm{in}}$ for modeling  with openEMS is assumed to be constant over the whole wavelength range.

\section*{Additional graphs}

\begin{figure}[h!]
{\centering
\includegraphics[width=\subfigsize]{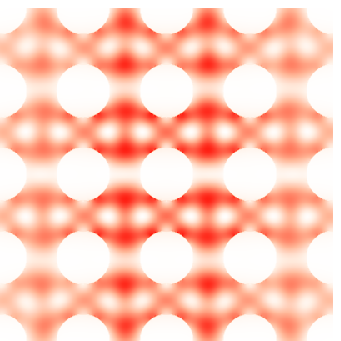}\hspace{1em}
\includegraphics[width=\subfigsize]{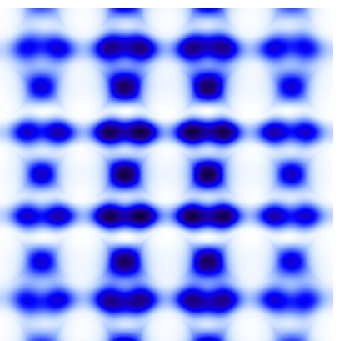}\par}
\caption{\label{ener_dist_GaP_361_lattcylhol} 
Time-averaged distributions of electric $|E|^2$ (left part, red color) and magnetic $|H|^2$ (right part, blue color) energy densities in the GaP nanohole array in proximity to the anapole state ($a=250$~nm, $h=100$~nm, $r=80$~nm, $\lambda=532$~nm).
The distributions are calculated within the plane $h/2$.
The incident Gaussian beam is polarized along the vertical direction.}
\end{figure} 

\begin{figure}[h!]
{\centering
\includegraphics[width=\figsize]{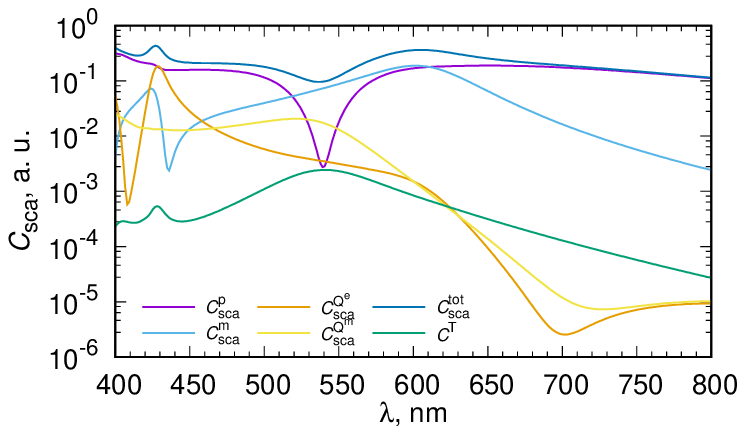}
\par} 
\caption{\label{Csca_exact_toroidalz_log_r080}
Scattering cross section spectra for the multipole contributions (electric dipole $C_\mathrm{sca}^\mathrm{p}$, magnetic dipole $C_\mathrm{sca}^\mathrm{m}$, electric quadrupole $C_\mathrm{sca}^\mathrm{Q^e}$, magnetic quadrupole $C_\mathrm{sca}^\mathrm{Q^m}$), their sum $C_\mathrm{sca}^\mathrm{tot}$ and the intensity of the electric dipole toroidal moment $C^\mathrm{T}$ for the GaP lattice element with $a=250$~nm, $h=100$~nm, $r=80$~nm. 
Refractive index $n_{0\,\mathrm{in}}$ is assumed to be constant over the whole wavelength range.
}
\end{figure} 

\begin{figure}%[h!]
{\centering
\includegraphics[width=\subfigsize]{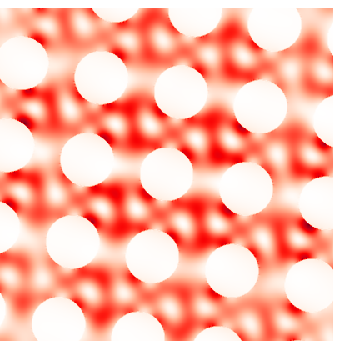}\hspace{1em}
\includegraphics[width=\subfigsize]{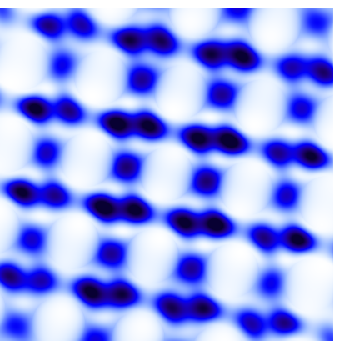}\par}
\vspace{1ex}
% \caption{\label{ener_dist_GaP_361_lattcylhol_rot10} 
% Time-averaged distributions of electric $|E|^2$ (left part, red color) and magnetic $|H|^2$ (right part, blue color) energy densities in the GaP nanohole array at the anapole mode ($a=250$~nm, $h=100$~nm, $r=80$~nm, $\lambda=532$~nm).
% The distributions are calculated within the plane $h/2$.
% The incident Gaussian beam is polarized along the vertical direction. The nanohole array is rotated by $10^\circ$ to the electric field polarization direction.}
% \end{figure} 
% 
% \begin{figure}%[h!]
{\centering
\includegraphics[width=\subfigsize]{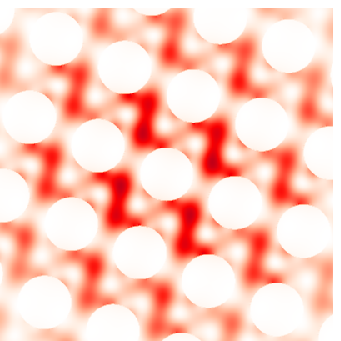}\hspace{1em}
\includegraphics[width=\subfigsize]{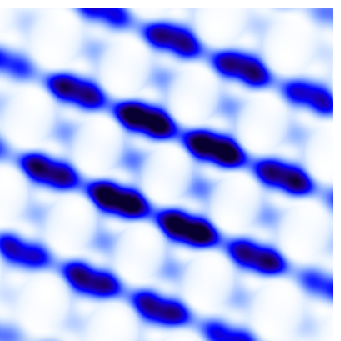}\par}
\caption{\label{ener_dist_GaP_361_lattcylhol_rot20} 
Time-averaged distributions of electric $|E|^2$ (left parts, red color) and magnetic $|H|^2$ (right parts, blue color) energy densities in the GaP nanohole array in proximity to the anapole state ($a=250$~nm, $h=100$~nm, $r=80$~nm, $\lambda=532$~nm).
The distributions are calculated within the plane $h/2$.
The incident Gaussian beam is polarized along the vertical direction. The nanohole arrays are rotated by $10^\circ$ (upper row) or $20^\circ$ (lower row) to the electric field polarization direction.}
\end{figure} 

\begin{figure}%[h!]
{\centering
\includegraphics[width=\figsize]{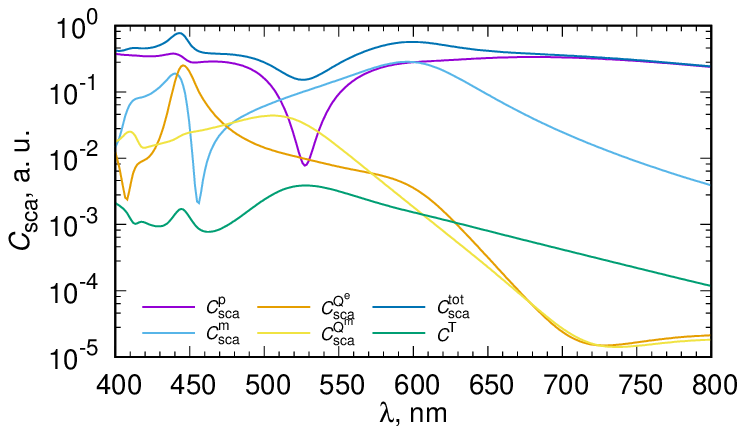}
\par} 
\caption{\label{Csca_exact_toroidalz_log_log_a300_r120}
Scattering cross section spectra for the multipole contributions (electric dipole $C_\mathrm{sca}^\mathrm{p}$, magnetic dipole $C_\mathrm{sca}^\mathrm{m}$, electric quadrupole $C_\mathrm{sca}^\mathrm{Q^e}$, magnetic quadrupole $C_\mathrm{sca}^\mathrm{Q^m}$), their sum $C_\mathrm{sca}^\mathrm{tot}$ and the intensity of the electric dipole toroidal moment $C^\mathrm{T}$ for the GaP lattice element with $a=300$~nm, $h=100$~nm, $r=120$~nm. 
Refractive index $n_{0\,\mathrm{in}}$ is assumed to be constant over the whole wavelength range.
}
\end{figure} 

\begin{figure}%[h!]
{\centering
\includegraphics[width=\figsize]{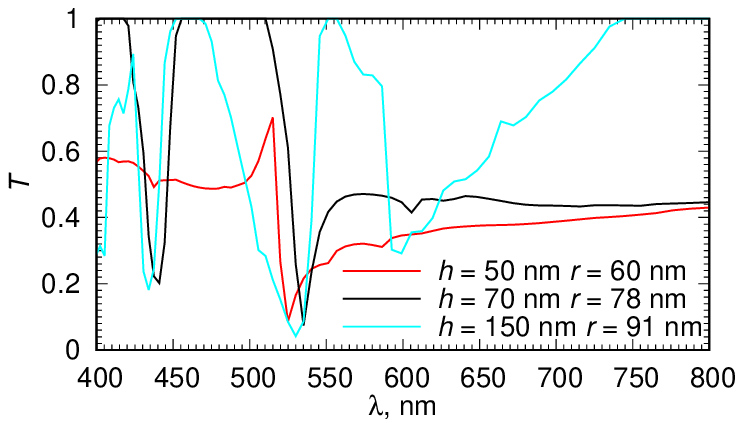}
\par} 
\caption{\label{Tspectr_GaP_h}
Transmission spectra for the GaP nanohole array with $a=250$~nm for various thicknesses in the vicinity of the anapole state. 
Refractive index $n_{0\,\mathrm{in}}$ is assumed to be constant over the whole wavelength range.
}
\end{figure} 

\begin{figure}%[h!]
{\centering
\includegraphics[width=\figsize]{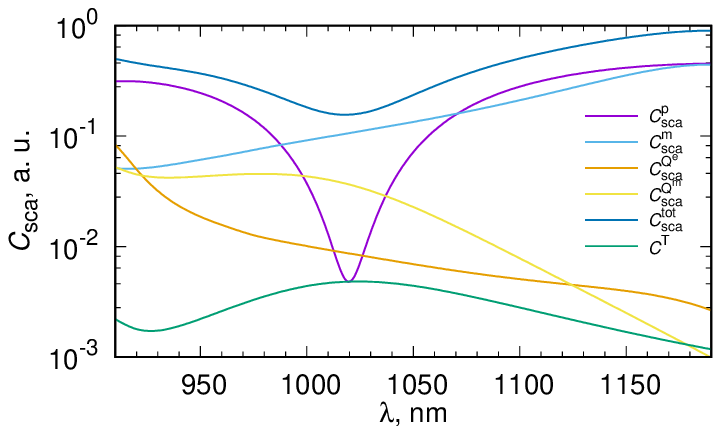}
\par} 
\caption{\label{Csca_Si_exact_toroidalz_log}
Scattering cross section spectra for the multipole contributions (electric dipole $C_\mathrm{sca}^\mathrm{p}$, magnetic dipole $C_\mathrm{sca}^\mathrm{m}$, electric quadrupole $C_\mathrm{sca}^\mathrm{Q^e}$, magnetic quadrupole $C_\mathrm{sca}^\mathrm{Q^m}$), their sum $C_\mathrm{sca}^\mathrm{tot}$ and the intensity of the electric dipole toroidal moment $C^\mathrm{T}$ for the Si lattice element with $a=500$~nm, $h=200$~nm, $r=180$~nm. 
Refractive index $n_{0\,\mathrm{in}}$ is assumed to be constant over the whole wavelength range.
}
\end{figure}

\begin{figure}[h!]
{\centering
\includegraphics[width=\figsize]{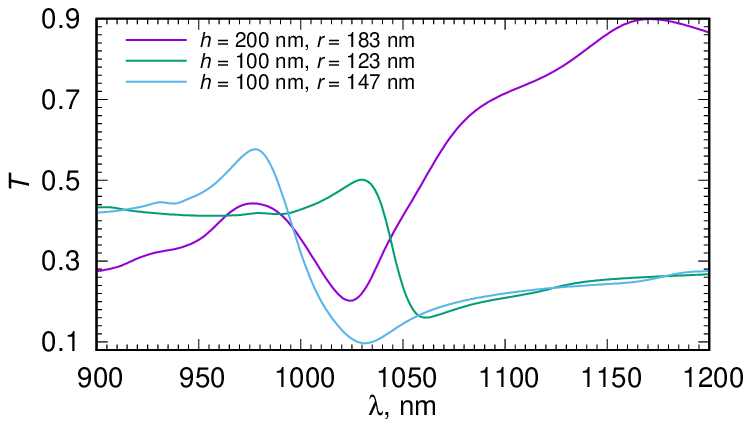}
\par} 
\caption{\label{Tspectr_Si}
Transmission spectrum for the Si nanohole array with $a=500$~nm. Refractive index $n_{0\,\mathrm{in}}$ is assumed to be constant over the whole wavelength range.
}
\end{figure}

%\bibliography{nlphase}

\end{document}